\documentclass[letterpaper]{IEEEtran}

\usepackage{flushend}

 \usepackage{graphicx}
\usepackage[cmex10]{amsmath}
\usepackage{amssymb}
\newtheorem{theorem}{Theorem}[section]
\newtheorem{lemma}{Lemma}[section]

\begin{document}
%
\title{An Approach using Demisubmartingales for the Stochastic Analysis of Networks}
%
%
%

\author{Kishore~Angrishi~and~Ulrich~Killat
\thanks{K. Angrishi is with the T-Systems International GmbH, 20146 Hamburg, Germany (e-mail: kishore.angrishi@t-systems.com).}
\thanks{U. Killat is with Hamburg University of Technology, 21073 Hamburg, Germany (e-mail: killat@tu-harburg.de).}}

\maketitle

\begin{abstract}
Stochastic network calculus is the probabilistic version of the  network calculus, which uses envelopes to perform probabilistic analysis of queueing networks. The accuracy of probabilistic end-to-end delay or backlog bounds computed using network calculus has always been a concern. In this paper, we propose novel end-to-end probabilistic bounds based on demisubmartingale inequalities which improve the existing bounds for the tandem networks of GI/GI/1 queues. In particular, we show that reasonably accurate bounds are achieved by comparing the new bounds with the existing results for a network of M/M/1 queues.
\end{abstract}


%
\IEEEpeerreviewmaketitle

\section{Introduction}
\IEEEPARstart{Q}{ueueing} theory is the mathematical study of queues, which generally uses probability mass or density functions to describe arrival traffic and service offered at the network node to compute probabilistic delay or backlog measures. However, with few exceptions, analysis of queueing networks to compute end-to-end probabilistic performance measures is mathematically complex without making simplifying assumptions on arrival traffic or service offered at the network nodes. In most situations, probabilistic bounds on performance measures are as sufficient as the actual values. Deterministic network calculus is an elegant theory, useful for computing worst-case bounds on end-to-end delay or backlog in queueing networks. Stochastic network calculus is the probabilistic extension of deterministic network calculus, which uses an envelope approach to describe arrival traffic and service offered at the network node. The tightness of the end-to-end probabilistic performance bounds has always been a concern in stochastic network calculus. The concern is mainly due to the use of union bounds for computing the bounds on probabilistic performance measures of the network. Recently, in \cite{florin:2007-1,florin:2007-2}, authors have derived new performance bounds for a GI/GI/1 queue in stochastic network calculus using Doob's maximal inequality for exponential supermartingales (instead of using union bounds) which are comparable to the exact results of M/M/1 and M/D/1 queues from queueing theory. A general comparison of results for GI/GI/1 queue from statistical network calculus with the classical queueing theory is made in \cite{jiang:2009}. 

In this paper, we compute end-to-end probabilistic performance bounds for tandem networks of GI/GI/1 queues in stochastic network calculus using demisubmartingale inequalities \cite{chris:2003,rao:2007}. The key difference of the approach used in this paper to the work presented in \cite{florin:2007-1,florin:2007-2} is that we derive performance bounds for a GI/GI/1 queue using statistical envelopes, in contrast to using stochastic processes as in \cite{florin:2007-1,florin:2007-2}. The rest of the paper is structured as follows: In Section \ref{sec:SNC}, we introduce the notion and assumptions used in the paper. In Section \ref{sec:bounds}, we derive the probabilistic end-to-end performance bounds on delay and backlog for the tandem networks of GI/GI/1 queues using statistical envelopes. Brief conclusions are presented in Section \ref{sec:conclusion}. 

\section{Notation and Assumptions}
\label{sec:SNC}
\begin{figure}
\centering
\includegraphics[scale=0.42]{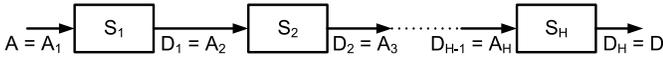}
\caption{ Network of H concatenated nodes}
\label{fig:tandemnet} 
\vspace{-5 mm}
\end{figure}

Our time model is discrete, i.e., $t \in \mathbb{N}_0 = \{0, 1, 2, \ldots \}$. We assume that the arrival traffic and the service offered at a node are stationary and have independent increments. In a network of nodes connected in series as shown in Fig. \ref{fig:tandemnet}, we use non-decreasing, left-continuous processes $A_h$ and $D_h$ to describe the arrivals and the departures at node $h$, respectively. $A_h(s,t)$ and $D_h(s,t)$ represent the cumulative amount of data seen in an interval $(s,t]$ at input and output of node $h$, respectively, for any $0 \le s \le t$. For the arrival and departure processes at node $h$, we assume the initial condition $A_h(t) = 0$ for $t \in (-\infty, 0]$ and the causal condition $D_h(t) \le A_h(t)$, where we denote $A_h(0,t) = A_h(t)$ and $D_h(0,t) = D_h(t)$ for any $t \ge 0$. The backlog $B_h(t)$ and delay $W_h(t)$ at time $t \ge 0$ in a node $h$ are given by $B_h(t) = A_h(t) - D_h(t)$ and $W_h(t) = \inf{\{d \ge 0: A_h(t-d) \le D_h(t)\}}$, respectively.

A stochastic process $S_h$ is said to describe the service offered at node $h$, if the corresponding arrival and departure processes at node $h$ satisfy for any fixed sample path and $t \ge 0$:
\begin{equation}
 A_h\otimes S_h(t) \le D_h(t)
 \label{reffsenv}
\end{equation}
where $\otimes$ is the min-plus convolution of $A_h$ and $S_h$ which is defined as $A_h\otimes S_h(t) = \inf_{0 \le u \le t} \{ A_h(0,u) + S_h(u,t)\}$. Any random process $S$ satisfying the above relationship is referred to as ``dynamic F-server'' in \cite{chang:2000}. 

The arrival and the service processes are described using statistical envelopes in network calculus. A statistical arrival envelope ${\cal G}$ for an arrival process $A$ is defined as a non-negative function for all $t \geq 0$ satisfying the following condition
\begin{equation}
	P\{A(t) - {\cal G}(t) > \sigma\} \le \varepsilon_{{\cal G}} (\sigma)
	 	\label{effenv}
\end{equation}
where $\varepsilon_{{\cal G}}$ is a non-increasing error function bounding the violation probability of the statistical arrival envelope. Similarly, a statistical service envelope ${\cal S}$ describing the service offered at the network node with arrival traffic $A$ and departure traffic $D$ is defined as a non-negative function for all $t \geq 0$ satisfying the following condition
\begin{equation}
	P\{A \otimes {\cal S}(t) - D(t) > \sigma\} \le \varepsilon_{{\cal S}} (\sigma)
	 	\label{effsenv1}
\end{equation}
where $\varepsilon_{{\cal S}}$ is a non-increasing error function bounding the violation probability of the statistical service envelope. The statistical service envelope from equation (\ref{effsenv1}) is related to the service process from equation (\ref{reffsenv}) for all $t \ge 0$ by the following expression
\begin{equation}
	P\{{\cal S}(t) - S(t) > \sigma\} \le \varepsilon_{{\cal S}} (\sigma)
	 	\label{effsenv}
\end{equation} 
In this paper, we use the notion of effective bandwidth ($\alpha$) \cite{kelly:1996} and effective capacity ($\beta$) \cite{kumar:2001,wu:2003,MMB:2008} from large deviations theory to derive statistical arrival and service envelopes describing the stochastic arrival traffic and the service offered at a node, respectively. The effective bandwidth of an arrival traffic $A$ with independent increments from \cite{kelly:1996}, for any $ \theta, t > 0$, is given as
\begin{equation}
\alpha(\theta) = \frac{1}{\theta} \log{E\left[ e^{\theta A(1)}\right]}  
\label{eb}
\end{equation}
Similarly, the effective capacity function of a stochastic service process $S$ with independent increments, for any $\theta, t > 0$, is defined as
\begin{equation}
{\beta}(\theta) = - \frac{1}{\theta} \log{E\left[ e^{-\theta  S(1)}\right]}
	 	\label{ec}
\end{equation} 
Then the statistical arrival and service envelopes in terms of effective bandwidth of the probabilistic arrival process and effective capacity of the service process observed at a network node are given as ${\cal G}(t) = \alpha(\theta)t$ and ${\cal S}(t) = \beta(\theta)t$, respectively, for any given $\theta \ge 0$; they satisfy the appropriate conditions in equations (\ref{effenv}) and (\ref{effsenv}) with the error function $\varepsilon (\sigma) = e^{-\theta \sigma}$.
The main advantage of using network calculus to do performance analysis of networks is that the network calculus allows to model a network of nodes as a single virtual node. The stochastic network service process $S_{net}$ characterizing the service offered in a single virtual network node, which represents a network of $H$ nodes connected in series as shown in Fig.\ref{fig:tandemnet}, can be computed for any fixed sample path using the min-plus convolution of the stochastic service process $S_h$ of constituting nodes for $h=1, \ldots, H$, i.e., $S_{net} = S_1 \otimes S_2 \otimes \cdots \otimes S_H$ \cite{chang:2000,fidler:2006}. The corresponding statistical network service envelope is given as ${\cal S}_{net} = {\cal S}_{1} \otimes {\cal S}_{2} \otimes \cdots \otimes {\cal S}_{H}$, where ${\cal S}_{h}$ is the statistical service envelope describing the service offered at node $h$, for $h = 1, 2, \ldots, H$. We assume that the arrival traffic $A_1$ at the ingress of the network and the stochastic service processes $S_h$, for $h=1, \ldots, H$, characterizing the service offered at the nodes of the network are independent of each other.

\section{Probabilistic Bounds on Backlog and Delay}
\label{sec:bounds}

In this section, we compute probabilistic bounds on backlog and delay in a network of $H$ nodes as shown in Fig. \ref{fig:tandemnet} using demisubmartingale inequalities. Let $A_1 = A$ and $D_H = D$ be the arrival traffic at the ingress of the network and departure traffic from the egress of the network, respectively.  
The following theorem provides the probabilistic bounds on end-to-end backlog and delay using the statistical envelopes of arrival and service processes at each network node $h$, respectively. 
\begin{theorem}
\label{theorem:pbr}
Let the service offered at node $h$ in a tandem network be characterized by the stochastic service process $S_h$ with the corresponding effective capacity function $\beta_h$, for $h=1, \ldots, H$. Let $A$ be the arrival process with effective bandwidth $\alpha$ and $D$ be the departure process from the tandem network with $H$ nodes. Then we have the following bounds.
\begin{enumerate}
	\item Backlog bound : The probabilistic bound on the backlog in a network, for any $t \ge 0$, is given by
				\begin{equation}
					P\left\{B(t) > x \right\} \le \tilde{\varepsilon}(x) \label{backlog}
				\end{equation}	
	\item Delay bound : The probabilistic bound on the delay in a network, for any $t \ge 0$, is given by
				\begin{equation}
					P\left\{W(t) > d \right\} \le \tilde{\varepsilon}(\alpha(\theta^*)d) \label{delay}
				\end{equation}
\end{enumerate}
where  $\tilde{\varepsilon}$ is an error function, for any $x \ge 0$, given as: 
\small	
\begin{equation}
	\tilde{\varepsilon}(x)=
	\begin{cases}
				\ \ \ \ \ \ \ \ \ \ \ \ \ \ \ \ \ \ \ \ \ e^{-\theta^*x} & \text{if } H=1\\
				 e^{-(\theta^*x-(H-1))}\sum_{h=0}^{H}{\frac{(\theta^*x-(H-1))^h}{h!}} & \text{if } H>1\\
				 & \text{and }  x \ge \frac{H-1}{\theta^*}
				\end{cases} \label{errorfn}
\end{equation} and $\theta^* = \sup{\{\theta : \alpha(\theta) \le \min_{1 \le h \le H}{\{\beta_h(\theta)\}}\}}$.
\normalsize
\end{theorem}
The proof of the theorem relies on applying demisubmartingale inequalities to compute probabilistic bounds. The key observation is that certain functions of the random arrival and service processes together with their corresponding statistical envelopes form demisubmartingales\footnote{A  sequence $\left\{S_n, n\ge 1\right\}$ is said  to  be  a  demisubmartingale  if $E\left[(S_{j+1}-S_j)f(S_1, \ldots, S_j)\right] \ge 0, j=1,2,\ldots$  for  every nonnegative  coordinatewise  nondecreasing function $f$ whenever the expectation is defined. If $E\left[(S_{j+1}-S_j)f(S_1, \ldots, S_j)\right] \le 0, j=1,2,\ldots$ sequence $\left\{S_n, n\ge 1\right\}$ is said  to  be  a  N-demisupermartingale}. This is shown using the following lemma.
\begin{lemma}
\label{lemma:demi}
Let $A$ be the arrival traffic with effective bandwidth $\alpha$ at a network node offering a stochastic service characterized by a service process $S$ with effective capacity $\beta$. If the arrival and the service processes have stationary independent increments, then the random processes $X(t) = e^{\theta (A(t) - \alpha(\theta)t)}$, $Y(t) = e^{-\theta (S(t) - \beta(\theta)t)}$, $Z(t) = e^{-\theta^* (S(t) - A(t))}$  and $Y^*(t) = \sup_{0 \le u \le t} Y(u)$ are demisubmartingales in an interval $[0,t]$ for $t \in \mathbb{N}_0 = \{0, 1, 2, \ldots \}$ and any $\theta \in (0,\theta^*]$, where $\theta^* = \sup{\{\theta : \alpha(\theta) \le \beta(\theta)\}}$. 
\end{lemma}
\textbf{\textit{Proof:}} To prove that $X(t)$, $Y(t)$, $Z(t)$ and $Y^*(t)$ are demisubmartingales \cite{chris:2003,rao:2007} for $t \in \mathbb{N}_0 = \{0, 1, 2, \ldots \}$ and any $\theta \in (0,\theta^*]$ , we need to show that $E[(X(t+1)-X(t))f(X(1), X(2), \ldots, X(t))] \ge 0$, and corresponding statements hold for $Y(t), Z(t)$ and $Y^*(t)$ for $t \in \mathbb{N}_0 = \{0, 1, 2, \ldots \}$ and every co-ordinatewise non-decreasing, non-negative function $f$ whenever the expectation is defined. As the proof for $X(t)$ follows the same lines as $Y(t)$, we will provide the proofs only for $Y(t)$, $Z(t)$ and $Y^*(t)$.
\small
\begin{eqnarray*}
\lefteqn{E[(Y(t+1) - Y(t))f(Y(1), Y(2), \ldots, Y(t))]} \\
&=& E[(e^{\theta (\beta(\theta) - S(t,t+1))} - 1)Y(t)f(Y(1), Y(2), \ldots, Y(t))]\\
&=& E[e^{\theta (\beta(\theta) - S(t,t+1))} - 1]E[g(Y(1), Y(2), \ldots, Y(t))] = 0
\end{eqnarray*}
\normalsize
The last two equalities are due to the fact that the process $Y(t)$ has independent increments and $E[e^{-\theta S(t,t+1)}] = e^{-\theta \beta(\theta)}$ (cf., equation (\ref{ec})), respectively.  This proves that $Y(t)$ is a demisubmartingale (also a N-demisupermartingale). 
\small
\begin{eqnarray*}
\lefteqn{E[(Z(t+1) - Z(t))f(Z(1), Z(2), \ldots, Z(t))]} \\
&=& E[(e^{-\theta^* (S(t,t+1)-A(t,t+1))} - 1)Z(t)f(Z(1), \ldots, Z(t))]\\
&=& (e^{-\theta^* \{\beta(\theta^*)-\alpha(\theta^*)\}} - 1)E[g(Z(1), \ldots, Z(t))] = 0
\end{eqnarray*}
\normalsize
Equality at the second step is due to our assumption that the arrival $A(t)$ and service $S(t)$  processes have independent increments and $E[e^{-\theta S(t,t+1)}] = e^{-\theta \beta(\theta)}$, $E[e^{\theta A(t,t+1)}] = e^{\theta \alpha(\theta)}$. The last equality is from stability condition \footnote{The stability condition for the queue at a node is $\alpha(\theta) \le \beta(\theta)$ for any finite $\theta \in (0,\infty)$.} and the definition of $\theta^*$. This proves that $Z(t)$ is a demisubmartingale (also a N-demisupermartingale).
\small 
\begin{eqnarray*}
\lefteqn{E[(Y^*(t+1) - Y^*(t))f(Y^*(1), Y^*(2), \ldots, Y^*(t))]} \\
&=& E[(\max{\{Y^*(t),e^{\theta (\beta(\theta) - S(t,t+1))}\}} - Y^*(t))\\
&& \ \ \ \ \ \ \ \ \ \ \ \ \ \ \ \ \ \ \ \ \ \ \ \ \ \ \ \ f(Y^*(1), Y^*(2), \ldots, Y^*(t))]\\
&=& E[\max{\{0,e^{\theta (\beta(\theta) - S(t,t+1))}-Y^*(t)\}} \\
&&  \ \ \ \ \ \ \ \ \ \ \ \ \ \ \ \ \ \ \ \ \ \ \ f(Y^*(1), Y^*(2), \ldots, Y^*(t))] \ge 0
\end{eqnarray*}
\normalsize
This proves that $Y^*(t)$ is a demisubmartingale.$\blacksquare$\\
By Doob's maximal inequality for demisubmartingales \cite{chris:2003,rao:2007}, we have the following maximal inequalities for any $\theta, \sigma \ge 0$,
\small
\begin{eqnarray}
P\left\{ \sup_{0 \le u \le t}X(u) > e^{\theta \sigma} \right\} &\le& E[X(t)]e^{-\theta \sigma} = e^{-\theta \sigma} \label{doobX}\\
P\left\{ \sup_{0 \le u \le t}Y(u) > e^{\theta \sigma} \right\} &\le& E[Y(t)]e^{-\theta \sigma} = e^{-\theta \sigma} \label{doobY}\\
P\left\{ \sup_{0 \le u \le t}Z(u) > e^{\theta^* \sigma} \right\} &\le& E[Z(t)]e^{-\theta^* \sigma} = e^{-\theta^* \sigma} \label{doobZ}\\
P\left\{ \sup_{0 \le v \le u \le t}Y(v,u) > e^{\theta \sigma} \right\} &=& P\left\{ \sup_{0 \le u \le t}Y^*(u) > e^{\theta \sigma} \right\} \nonumber \\
&\le& E[Y^*(t)]e^{-\theta \sigma} \nonumber \\
&\le& eE[Y(t)]e^{-\theta \sigma} = ee^{-\theta \sigma} \label{doobY*}
\end{eqnarray}
\normalsize
The final inequality step is due to Rao's maximal inequality for demisubmartingales (Theorem $3.7$ from \cite{rao:2007}). The proof of Theorem \ref{theorem:pbr} also relies on Lemma 4.1 from \cite{yuming:2006}, which states that for any two non-negative independent random variables $F$ and $G$ with $P(F > \sigma) \le f(\sigma)$ and $P(G > \sigma) \le g(\sigma)$ where $f(\sigma)$ and $g(\sigma)$ are non-negative, decreasing functions for any $\sigma \ge 0$, then 
\begin{equation}
P\left\{F + G > \sigma \right\} \le 1-\int_{0}^{\sigma}{\tilde{f}(\sigma-u)d\tilde{g}(u)}  
\label{speffenv1} 
\end{equation}
where $\tilde{f}(\sigma) =  1 - \left[f(\sigma) \right]^- $, $\tilde{g}(\sigma) = 1 - \left[g(\sigma)\right]^-$ and $[a]^- = \min(1,a)$ for any $a \ge 0$.\\
\textbf{\textit{Proof of Theorem \ref{theorem:pbr}:}} We now provide the proof for the probabilistic end-to-end delay bound. The proof for the probabilistic bound on end-to-end backlog is its immediate variation. For single hop ($H=1$), the proof is straight forward and can be shown for fixed sample path, $t \ge 0$ and $\theta > 0$ as follows:
\small
\begin{eqnarray}
\lefteqn{P\left\{ W(t) > d \right\} = P\left\{ A(t-d) - D(t) > 0 \right\}}  \nonumber\\
&=& P\left\{ A(t-d) - D(t) > 0 \right\} \nonumber\\
&\le& P\left\{ A(t-d) - A \otimes S_1(t)  > 0  \right\} \nonumber\\
&=& P\left\{\sup_{0 \le u \le t-d}\{A(u,t-d) - S_1(u,t)\} + \alpha(\theta^*)d > \alpha(\theta^*)d \right\} \nonumber\\ 
&=&P\left\{\sup_{0 \le u \le t-d}\left\{e^{\theta^*\{A(u,t-d) - S_1(u,t) + \alpha(\theta^*)d \}} \right\} > e^{\theta^*\alpha(\theta^*)d} \right\} \nonumber\\
&\le&E\left[ e^{\theta^*\{A(0,t-d) - S_1(0,t-d) - S_1(t-d,t) + \alpha(\theta^*)d} \right] e^{-\theta^*\alpha(\theta^*)d}\nonumber\\
&=&e^{\theta^* \{ \alpha(\theta^*)(t-d) - \beta_1(\theta^*)(t-d) - \beta_1(\theta^*)d + \alpha(\theta^*)d\}}e^{-\theta^* \alpha(\theta^*)d}\nonumber\\
&=&e^{-\theta^* \alpha(\theta^*)d} \label{s1d}
\end{eqnarray}
\normalsize
The first inequality is from the definition of stochastic network service process from equation (\ref{reffsenv}). The final inequality is due to Doob's inequality for demisubmartingales from equation (\ref{doobZ}). The last two steps are due to our assumption that the arrival $A(t)$ and service $S(t)$  processes have independent increments and due to the stability condition, respectively. For $H>1$ and for fixed sample path, $t \ge 0$ and $\theta > 0$, we have,
\small
\begin{eqnarray} 
\lefteqn{P\left\{ W(t) > d \right\} = P\left\{ A(t-d) - D(t) > 0 \right\}} \nonumber\\
&\le& P\left\{ A(t-d) - A\otimes S_{net}(t) > 0 \right\} \nonumber\\
&=& P\left\{ A(t-d) - A\otimes S_1\otimes S_2 \otimes \cdots \otimes S_H(t) > 0 \right\}\\
&=& P\left\{\sup_{0 \le k_1 \le k_2 \le k_3 \le \cdots \le k_H \le t} \{A(t-d) - A (k_1) \right. \nonumber\\
&& \ \ \ \ \ \ \ \ \ \ \left. -  S_1(k_1,k_2)  - S_2(k_2,k_3) - \cdots - S_H (k_H,t) \} \right. \nonumber\\
&& \ \ \ \ \ \ \ \ \ \ \left.  + \sup_{0 \le u}\{{\cal G}(u-d)-{\cal S}_{1}\otimes{\cal S}_{2}\otimes \cdots \otimes {\cal S}_{H}(u)\} \right. \nonumber\\
&& \ \ \ \ \ \ \ \ \ \ \ \ \ \ \ \left. > \sup_{0 \le u}\{{\cal G}(u-d)-{\cal S}_{1}\otimes{\cal S}_{2}\otimes \cdots \otimes {\cal S}_{H}(u)\} \right\}\nonumber\\ 
&\le& P\left\{\sup_{0 \le k_1 \le k_2 \le k_3 \le \cdots \le k_H \le t} \{A(k_1,t-d)  \right. \nonumber\\
&&  \ \ \ \ \ \ \ \ \left. - \alpha(\theta)(t-k_1-d) + \beta_1(\theta)(k_2-k_1) -  S_1(k_1,k_2) \right. \nonumber\\
&&  \ \ \ \ \ \ \ \ \ \ \ \ \ \ \left. + \beta_2(\theta)(k_3-k_2)  - S_2(k_2,k_3) + \cdots  \right. \nonumber\\
&&  \ \ \ \ \ \ \ \ \ \ \ \ \ \ \ \ \ \ \ \ \left. + \beta_H(\theta)(t-k_H) - S_H (k_H,t) \}  > \alpha(\theta)d \right\} \nonumber\\
&\le& P\left\{ \sup_{0 \le k_1 \le k_2 \le t}\{\beta_1(\theta)(k_2-k_1)- S_1(k_1,k_2)\} \right. \nonumber\\
&& \ \ \ \ \ \ \ \ \left. + \sup_{0 \le k_2 \le k_3 \le t} \{\beta_2(\theta)(k_3-k_2)- S_2(k_2, k_3)\} + \cdots \right. \nonumber\\
&& \ \ \ \ \ \ \ \ \ \ \left. + \sup_{0 \le k_H \le t} \{ \beta_H(\theta)(t-k_H) - S_H(k_H, t)\} \right. \nonumber\\
&& \ \ \ \ \left. + \sup_{0 \le k_1 \le t} \{A(k_1,t-d) + \alpha(\theta)(t-k_1-d) \} > \alpha(\theta)d \right\} \nonumber\\
&\le& e^{-(\theta^*\alpha(\theta^*)d-(H-1))}\sum_{h=0}^{H}{\frac{(\theta^*\alpha(\theta^*)d-(H-1))^h}{h!}} \label{snetd}
\end{eqnarray}
\normalsize
The first inequality is from the definition of stochastic network service process from equation (\ref{reffsenv}). We set ${\cal G}(t) = \alpha(\theta)t$ and ${\cal S}_h(t) = \beta_h(\theta)t$, for $h = 1, 2, \ldots, H$, with the stability condition $\alpha(\theta) \le \min_{1\le h \le H }{\{ \beta_h(\theta)\}}$ for any $\theta \ge 0$ and the justification for this network stability condition lies in the (approximate) invariance of the effective bandwidth $\alpha(\theta)$ \cite{spects:2008}. After some reordering we obtain the second inequality. The third inequality is from a property of supremum operation \footnote{$ \sup_{0\le s \le t} \{X(s) + Y(s)\}$ $ \le $ $\sup_{0\le s \le t} \{X(s)\}$ $+$ $\sup_{0\le s \le t} \{Y(s)\}$} \cite{yuming:2006}. We get the final inequality from Lemma \ref{lemma:demi}, equations (\ref{doobX}), (\ref{doobY}), (\ref{doobY*}) and (\ref{speffenv1}). The proof is obtained by a complete induction over $H$. The final expression is valid only for $\alpha(\theta^*)d \ge \frac{H-1}{\theta^*}$. This constraint is a consequence of the $[a]^-$ operation in equation (\ref{speffenv1}).$\blacksquare$\\
It can be observed that setting $H=1$ in equation(\ref{snetd}), one gets the probability bound to be $(1+\theta^*\alpha(\theta^*)d)e^{-\theta^*\alpha(\theta^*)d}$. Though the bound is valid, it can be easily verified that it is worse than the bound $e^{-\theta^*\alpha(\theta^*)d}$ from equation (\ref{s1d}). This discrepancy between the two bounds is due to the fact that for the bound from equation (\ref{s1d}) the stochastic process $\sup_{0\le u \le t-d}\{A(u,t-d)-S_1(u,t)\}$ is directly used to determine the delay bound, in contrast to the bound from equation (\ref{snetd}) where the arrival process with statistical arrival envelope and service process with service envelope are used individually to establish the result. 
\begin{figure}
\centering
\includegraphics[scale=0.31,angle=-90]{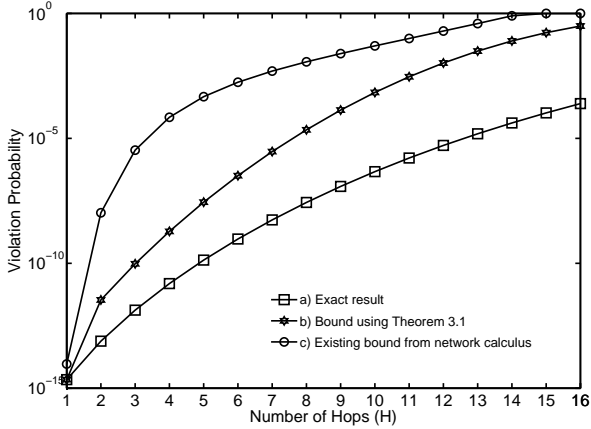}
\caption{Violation probability of the end-to-end delay bound ($\mu \cdot d=112.5$) in an M/M/1 network as a function of the number of
nodes; utilization factor at each node is $\rho = 0.7$}
\label{fig:demiDelay} 
\vspace{-5 mm}
\end{figure}

To analyse the accuracy of the new probabilistic end-to-end delay bound from Theorem \ref{theorem:pbr}, we compare it with the existing probabilistic bounds from network calculus and results from queueing theory for a network of M/M/1 queues. In an M/M/1 queuing system with one server, both the arrival and the service processes are of Poisson type. The customers arrive at rate $\lambda$ and the server works at rate $\mu$. We denote the utilization factor by $\rho = \lambda / \mu $, and assume for stability that $\rho < 1$. The effective bandwidth and effective capacity of the arrival and service processes (Poisson process) are $\lambda \frac{e^{\theta}-1}{\theta}$ and $\mu \frac{1 - e^{-\theta}}{\theta}$, respectively, with $\theta^* = -\log{\rho}$. We consider a special case of the network from Fig. \ref{fig:tandemnet} with $H$ M/M/1 queues connected in series to analyse the accuracy of our end-to-end network calculus delay bound. A Poisson flow with rate $\lambda$ traverses through the entire network. The arrival process at the downstream queue is the departure process of the upstream queue which is again Poissonian. Let each queue in the network be served by a similar service process $S$ with effective capacity $\beta(\theta)$ and let the service processes at all nodes of the network be independent of each other. It is known from queueing theory that the exact distribution of steady state end-to-end delay of the through flow $W(t)$ in a M/M/1 queueing network is given by 
\small
\begin{equation}
P\left\{W(t) > d \right\} =  \sum_{h=0}^{H-1} \frac{(\mu (1-\rho)d)^h}{h!} e^{-\mu (1-\rho)d} \label{delayNetQueue}
\end{equation}
\normalsize
The equation is obtained from an $H$-fold convolution of the (exponential) probability function of delay for a single M/M/1 node, followed by an integration in the limits from d to infinity.
The best available end-to-end delay bound of the through flow $W(t)$ in a M/M/1 queueing network from network calculus \cite{florin:2007-1,fidler:2006} is given by
\small
\begin{equation}
P\left\{W(t) > d \right\} =  \inf_{0 \le \theta \le \theta^*}{\left(\frac{1}{1-e^{-\theta\{\beta(\theta)-\alpha(\theta)\}}}\right)^H e^{-\theta \alpha(\theta)d}} \label{delayNetSNC}
\end{equation}
\normalsize
In Fig. \ref{fig:demiDelay}, we illustrate the violation probability of the delay bounds ($\mu \cdot d=112.5$) from equations (\ref{delay}) and (\ref{errorfn}) as curve (b) along with the exact results from equation (\ref{delayNetQueue}) as curve (a) and existing delay bounds from stochastic network calculus using moment generating functions \cite{fidler:2006,florin:2007-1} from equation (\ref{delayNetSNC}) as curve (c) for a fixed utilization factor $\rho = 0.7$ at each of the $H$ queues. It can be observed that the new results provide better bounds than the existing bounds and also follow the shape of the exact results from queueing theory.

\section{Conclusions}
\label{sec:conclusion}
In this paper we used demisubmartingale inequalities to compute end-to-end probabilistic delay and backlog bounds within the framework of network calculus. The tightness of the computed end-to-end probabilistic performance bounds is explored by comparing new bounds with the exact results from queueing theory for a network of M/M/1 queues.
\bibliographystyle{IEEEtran}
\bibliography{biblio}
%
%








\end{document}